\documentclass[%
 reprint,
 amsmath,amssymb,
 aps,
]{revtex4-1}
\usepackage{mathrsfs} 
\usepackage{fancybox}
\usepackage[utf8]{inputenc}
\usepackage{tikz}
\usetikzlibrary{patterns,decorations.pathreplacing}
\usetikzlibrary{calc}
\usepackage{graphicx}
\usepackage{dcolumn}
\usepackage{bm}

\begin{document}

\preprint{APS/123-QED}

\title{Fresnel coefficients and Fabry-Perot formula for spatially dispersive metallic layers}

\author{Armel Pitelet}
\author{\'Emilien Mallet}
\author{Emmanuel Centeno}
\author{Antoine Moreau}
 \email{antoine.moreau@uca.fr}
\affiliation{%
 Universit\'e Clermont Auvergne, CNRS, Institut Pascal, 63000 Clermont-Ferrand, France
}%




\date{\today}


\begin{abstract}
The repulsion between free electrons inside a metal makes its optical response spatially dispersive, so that it is not described by Drude's model but by a hydrodynamic model. We give here fully analytic results for a metallic slab in this framework, thanks to a two-modes cavity formalism leading to a Fabry-Perot formula, and show that a simplification can be made that preserves the accuracy of the results while allowing much simpler analytic expressions. For metallic layers thicker than 2.7 nm modified Fresnel coefficients can actually be used to accurately predict the response of any multilayer with spatially dispersive metals (for reflection, transmission or the guided modes). Finally, this explains why adding a small dielectric layer[Y. Luo {\em et al.}, Phys. Rev. Lett. 111, 093901 (2013)] allows to reproduce the effects of nonlocality in many cases, and especially for multilayers.
\end{abstract}

\pacs{Valid PACS appear here}
\maketitle


Drude's model, where the electromagnetic response of metals is summarized in a local permittivity, has been very successful in describing the optical response of metals even at the scale of a few nanometers. Electrons are however repulsing each other, making the response of metals spatially dispersive - a phenomenon that is completely overlooked in Drude's model. The response is then said to be non-local because the metal can not be described by a simple permittivity any more. This subject has attracted a lot of interest from a theoretical point of view in the seventies and eighties\cite{boardman82,forstmann86}, but an experimental evidence that the Drude model could be inaccurate even in the optical domain has been produced only very recently for very narrow gaps between two metals\cite{ciraci12,ciraci14}. The hydrodynamic model with hard-wall boundaries\cite{raza11,moreau13,ciraci13} is a sufficiently accurate framework to take these nonlocal effects into account - even if more complex models taking spill-out corrections have very rencently been proposed\cite{ciraci16}. It appears now that nonlocal effects have an impact on  metallo-dielectric multilayers with deeply subwavelength thicknesses of dielectric or metal\cite{yan12} for instance when guided modes are supported\cite{moreau13,raza13} or when trying to design all kinds of plasmonic flat lenses\cite{yan13,ruppin05,ruppin05b}. The hydrodynamic model is particularly interesting in the framework of multilayers because the fields have analytic expressions in that case\cite{ruppin05,moreau13,raza13,benedicto15,dechaux16}. Taking nonlocality into account can be complicated for more complex structures, and there is clearly a need for simpler approaches: it has been recently shown\cite{luo13}, spurring debate\cite{schaich_comment2015,luo_respcomment_2015}, that adding a very thin dielectric layer could yield results that match very well with the prediction of the hydrodynamic model.

\begin{figure}\centering
\begin{tikzpicture}[transform shape, scale=1]
\coordinate (x) at (7.5,0);
\coordinate (y) at (0,2);
\coordinate (hg) at (3,4);
\coordinate (bg) at ($ (hg)-(y) $);
\coordinate (hd) at ($ (hg)+(x) $);
\coordinate (bd) at ($ (hg)+(x)-(y) $);			\draw[gray!40,fill=gray!40] (hg) rectangle (bd);
\coordinate (mid) at ($ (hg)!0.5!(bd) $);
\coordinate (Origine) at ($ (hg)-0.5*(y) $);
\draw [->] (Origine) -- ++(0,0.75) node[pos=0.9, left]{$z$}; 
\draw [->] (Origine) -- ++(0.75,0) node[pos=0.9, below]{$x$};
\draw[dashed] (hg) -- ++(-0.25,0) node[left]{$0$};
\draw (hg) -- (hd);
\draw[dashed] (hd) -- ++(0.25,0);
\draw[dashed] (bg) -- ++(-0.25,0) node[left]{$-h$}; 
\draw (bg) -- (bd);
\draw[dashed] (bd) -- ++(0.25,0);
\node at ($ (mid)-0.3*(x) $) {$Metal$};
\node at ($ (mid)-0.3*(x)+0.75*(y) $) {$Insulator$};
\node at ($ (mid)-0.3*(x)-0.75*(y) $) {$Insulator$};
\node at ($ (mid)+0.75*(y) $) {$(1)$};
\node at ($ (mid) $) {(m)};
\node at ($ (mid)-0.75*(y) $) {$(2)$};
\node at ($ (mid)+0.3*(x)+0.75*(y) $) {$A_{1}e^{\kappa_{1}z}+B_{1}e^{-\kappa_{1}z}$};
\node at ($ (mid)+0.3*(x) $) {$A_{m}e^{\kappa_{m}z}+B_{m}e^{-\kappa_{m}z}$};
\node at ($ (mid)+0.3*(x)-0.75*(y) $) {$A_{2}e^{\kappa_{2}z}$};
\end{tikzpicture}
\caption{Schematic representation of an IMI slab of thickness $h$. The general expressions for the magnetic field $H_y$ are given for each layer, when light is assumed to illuminate the structure from above.}
\label{system}
\end{figure}
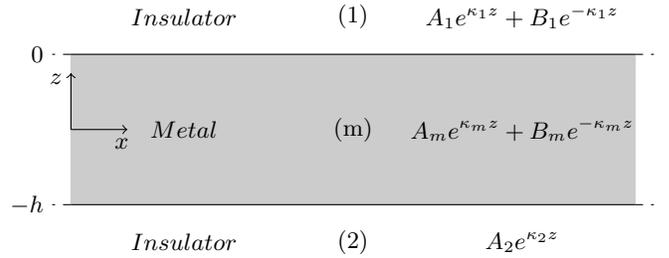

In the present work, we first obtain simple analytic expressions using a generalized cavity formalism for a single metallic slab. We then show  that a simple assumption, which is valid as soon as the metallic layer is thicker than 2.7 nm in the visible range and 5-6 nm in the near UV, can greatly reduce the complexity of the calculus of the nonlocal response. Using our assumption, simpler yet very accurate analytic results are obtained  for any kind of metallo-dielectric multilayer. In order to illustrate what this simplified model can bring and to clearly assess its potential, we study how nonlocality influences the guided modes of an insulator-metal-insulator (IMI) waveguide\cite{tournois97}, shown on Fig. \ref{system}. We show that nonlocality has a clear influence on the way these modes behave, especially when the metal in embedded in high index dielectric material. Our approach can be considered, in the framework of multilayered structures, as a justification for the work of Luo {\em et al.}\cite{luo13} because it is based on the use of effective Fresnel coefficients. Adding a very thin dielectric layer on top of the metal is actually a way to correct the reflection coefficients and thus to take nonlocality into account.


A metallic layer, when spatial dispersion is taken into account\cite{moreau13}, supports (i) the usual electromagnetic transverse wave and (ii) a longitudinal wave called bulk plasmon. This makes it possible to use a two-mode formalism to describe the optical response of the layer, even though both modes are evanescent. Such a formalism is usually utilized in the framework of resonant cavities\cite{sauvan05,dechaux16}  but here it provides analytic expressions for the reflection or the transmission of a metallic slab (see Fig. \ref{system}), which is not an easy task\cite{ruppin05}. Assuming an $e^{-i\omega t}$ time dependency,  these coefficients are obtained by solving the following system of equations\cite{dechaux16}, which is derived from the continuity equations:  
\begin{align}	
B_1= & r_{1m} A_1 + t_{m1} B_m + t_{m1}' B_l 	\label{reflexionnonlocal1}\\
A_m= & t_{1m} A_1 + r_{m1} B_m + r_{m1}' B_l 	\label{reflexionnonlocal2}\\
A_l= & \tau_{1m} A_1 + \rho_{m1} B_m + \rho_{m1}' B_l 	\label{reflexionnonlocal3}\\
B_m= & r_{m2} A_m e^{-2\kappa_mh} + r_{m2}' A_l e^{-(\kappa_m+\kappa_l)h}  \label{reflexionnonlocal4}\\
B_l= & \rho_{m2} A_m e^{-(\kappa_m+\kappa_l)h} + \rho_{m2}' A_l e^{-2\kappa_lh} \label{reflexionnonlocal5}
\end{align}  where $A_{(i)}$, $B_{(i)}$, ($i=1,2,m$ - see Fig. \ref{system}), are the amplitudes of transverse mode and $A_l$, $B_l$ are amplitudes of the longitudinal mode.  The above reflection (r,r',$\rho$,$\rho$') and transmission (t,t',$\tau$) coefficients are calculated using the boundary conditions following \cite{benedicto15} - the regular conditions and the additional boundary condition expressing that no electron is allowed to leave the metal. The $r$ and $t$ coefficients are respectively the reflection and transmission coefficients of the transverse modes, while $\rho'$ is the reflection coefficient for the longitudinal waves. When a longitudinal or a transverse wave is reflected by an interface, the other kind of wave is excited too. Coefficients $\rho$ and $r'$ take into account the conversion from transverse to longitudinal and the opposite respectively. Finally, $t'$ is the contribution to the outside plane wave from the longitudinal wave inside the metal while $\tau$ is the reverse. 

The attenuation constants $\kappa_{(i)}=\sqrt{k_{x}^2-\epsilon_{i}k_{0}^2}$ (with $k_0=\frac{2\pi}{\lambda}$) and $\kappa_l$ can be deduced from the dispersion relations of the transverse and longitudinal waves\cite{moreau13}, taking $k_x$ as the wavevector along the $x$ direction for all waves. We have thus 
\begin{equation}
 \kappa_l=\sqrt{{k_x}^2+\frac{{\omega_p}^2}{{\beta}^2}\left(\frac{1}{\chi_f}+\frac{1}{1+\chi_b}	\right)	}.
    \label{kappal}
\end{equation}
were $\chi_f = -\epsilon_0 \frac{{\omega_p}^2}{\omega^2+i\gamma\omega}$, $\chi_b$ are the susceptibilities of the free and bound electrons respectively that can be found in\cite{rakic98} for any metal, $\gamma=0.049$ eV and $\omega_p=8.16$ eV are respectively the damping factor and the plasma frequency for silver, and $\beta$ the non-local parameter, taking into account Coulomb interaction and the exchange interaction. This parameter is estimated from experimental data to be around $1.35.10^6$ m/s\cite{ciraci12,ciraci14}, quite close to theoretical predictions.

When the system describing the two-modes cavity is solved, it yields a generalized Fabry-Perot formula for the reflection coefficient, which is remarkable and very convenient\cite{dechaux16}. In the case of our transverse and longitudinal modes, the reflection coefficient is
\begin{equation}
r_{nl}=r_{eff}+\frac{\overline{r}_{m2} t_{eff} e^{-2\kappa_mh}}{1-\overline{\overline{r}}_{m1}\overline{r}_{m2}e^{-2\kappa_mh}}		\label{rnonlocal}
\end{equation}
where $r_{eff}$, $\overline{r}_{m2}$, $\overline{\overline{r}}_{m1}$, $t_{eff}$, are {\em effective} coefficients  based on the transmission and reflection coefficients of the two-modes formalism given above. Their exact expression, that is not needed here, can be found in \cite{dechaux16} and we underline that the effective reflection coefficients all still depend on $h$. Finally, since guided modes of such a structure correspond to poles of the reflection coefficient, the dispersion relation can simply be written $1-\overline{\overline{r}}_{m1}\overline{r}_{m2}e^{-2\kappa_mh}=0.$



The typical penetration length of the transverse wave is classically given by $\frac{1}{2\Re (\kappa_m)}$, which defines a power mode attenuation of $-4.34$ dB. For the longitudinal wave, this typical penetration length is roughly two orders of magnitude shorter and is given by $\frac{1}{2\Re(\kappa_l)}$. The large difference between the two penetration depth suggests it is possible to simplify all the above analytic expressions when the right conditions are met. 

However, at such typical depth into the metal, the field is still 37\% of the field at the interface - which is far from negligible. In order to get a more relevant penetration depth for the longitudinal wave, we define the quantity $L_{nl}$ as the distance inside the metal away from a single interface for which the field undergo a power mode attenuation of $-20$dB ($e^{-2\kappa_l L_{nl}}=10^{-2}$). Considering equations \eqref{reflexionnonlocal1} to \eqref{reflexionnonlocal5}, it is easy to understand that the slab response will be largely different whether it is thicker than $L_{nl}$ or not. If it is thicker, then all the terms in $e^{-\kappa_l h}$ can be neglected which leads to a great simplification: essentially, nonlocality only has an impact on the internal reflection/transmission coefficients of the transverse mode, which is the only wave that can eventually tunnel through the metallic slab. Considering this, we will refer to this approximation as the One Mode Approximation (1-MA). Another way to put this, is to say that the effective reflection coefficients of the exact solution depend on $h$ and tend to the 1-MA reflection coefficients when $h$ increases leading to the following reflection coefficient
\begin{equation}	r=r_{1m}+\frac{r_{m2}t_{m1}t_{1m} e^{-2\kappa_mh}}{1-r_{m1}r_{m2}e^{-2\kappa_mh}},
\end{equation}
where the expressions of $r_{ij}$ and $t_{ij}$ ($i,j=1,2,m$) reduce to 
\begin{align}		
  r_{ij}=\frac{\frac{\kappa_i}{\epsilon_i}-\frac{\kappa_j}{\epsilon_j}+\Omega}{\frac{\kappa_i}{\epsilon_i}+\frac{\kappa_j}{\epsilon_j}-\Omega} \label{rij}\\
t_{ij}=\frac{2\frac{\kappa_i}{\epsilon_i}}{\frac{\kappa_i}{\epsilon_i}+\frac{\kappa_j}{\epsilon_j}-\Omega}, \label{tij}
\end{align}
which constitute non-local Fresnel coefficients, and where
\begin{equation}
\Omega=\frac{{k_x}^2}{\kappa_l}\left(\frac{1}{\epsilon}-\frac{1}{1+\chi_b}\right).		\label{Omega}
\end{equation}

Finally, only the reflection coefficients of the Fabry-Perot like formulas are modified by nonlocality and this is enough to take the spatial dispersion into account as long as the metallic layer is thicker than a few nanometers. This largely simplifies the analytic calculations that can be made for any metallo-dielectric multilayer, especially for dispersion relations. Furthermore, this opens the possibility of computing the nonlocal response of any complex multilayer\cite{xu13} using a scattering matrix with slightly modified coefficients to take nonlocality into account, instead of a more complex method\cite{benedicto15}. When considering the interface between to media numbered $i$ and $j$, the outgoing waves with amplitudes $B_j$ and $A_i$ propagating downwards (respectively upwards) in the above (resp. lower) medium are linked to the incoming waves $A_i$ and $B_j$, propagating upwards (resp. downwards) in the upper (resp. lower) medium by a scattering matrix  $S_{i\rightarrow j}$ 
\begin{equation}
\begin{pmatrix} B_i \\ A_j \end{pmatrix} = S_{i\rightarrow j} \begin{pmatrix} A_i \\ B_j \end{pmatrix}= \begin{pmatrix} r_{ij} & t_{ji} \\ t_{ij} & r_{ji} \end{pmatrix} \begin{pmatrix} A_i \\ B_j \end{pmatrix}
\end{equation}
where the $r_{ij}$ and $t_{ij}$ are given by \eqref{rij} and \eqref{tij}. The rest of the scattering matrix method can be applied as it is\cite{defrance2016moosh}, it is not modified by the spatial dispersion.

With the 1MA, the dispersion relation for a metallic film is simply 
$1-r_{m1}r_{m2}e^{-2\kappa_mh}=0$.	
When the dielectric is the same above and below, the left part of the dispersion relation can be factorize to reach dispersion relations for symmetric and antisymmetric modes (when considering the transverse magnetic field). In the case of a metallic layer, each equation has only one solution. The symmetric mode is called the Long Range Surface Plasmon (LRSP)\cite{berini9} and its dispersion relation is 
\begin{align}		
\frac{\kappa_m}{\epsilon_m}\tanh{\left(\frac{\kappa_m h}{2}\right)}+\frac{\kappa_d}{\epsilon_d}-\Omega=0,
\end{align}\label{LRSP}
while the antisymmetric mode is the Short Range Surface Plasmon (SRSP) with a dispersion relation
\begin{align}
\frac{\kappa_m}{\epsilon_m}\coth{\left(\frac{\kappa_m h}{2}\right)}+\frac{\kappa_d}{\epsilon_d}-\Omega=0.
\end{align} \label{SRSP}

\begin{figure}\centering
\input{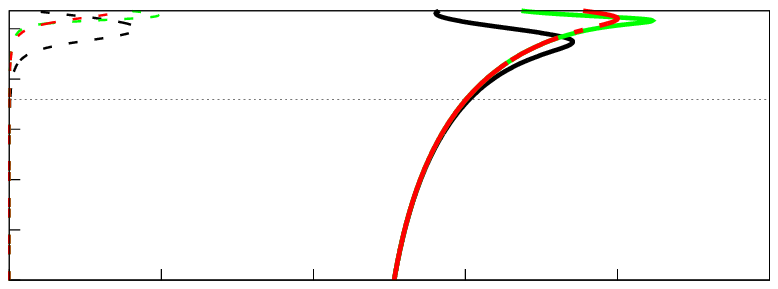}
\input{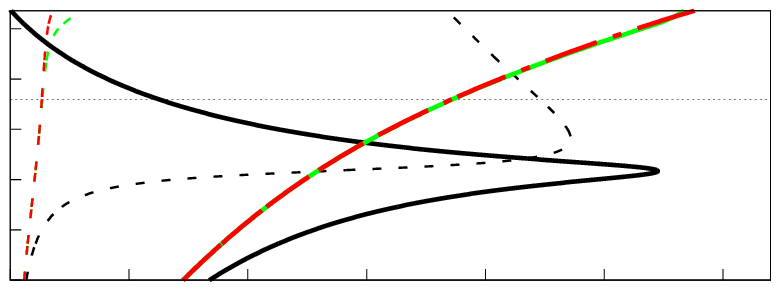}
\caption{Dispersion curves for the $TiO_2/Ag/TiO_2$ waveguide, showing the effective index $n_{eff}=k_x/k_0$ as a function of the normalized frequency $\frac{\omega}{\omega_p}$. The metal thickness $h$ is 1.7 nm. \textbf{(a)} shows the LRSP mode while \textbf{(b)} presents the SRSP mode. Red lines correspond to the exact solution (no assumption), green to the 1-MA (one-mode approximation) and black lines to Drude's model (local solution). Thick lines correspond to the real part of the effective index $Re(n_{eff})$, while thin lines show $Im(n_{eff})$.}
\label{dispersionTiO2}
\end{figure}

Guided modes supported by the IMI structure are well known for their high wavevectors. Since $\Omega$ increases with $k_x$, and since the impact of nonlocality is directly linked to $\Omega$, this leads us to expect an important impact of nonlocality on the guided modes supported by the IMI structure. That is the reason why in the following, we will use these guided modes to assess the impact of nonlocality.

Using a 1.7 nm (12 atoms in thickness) of silver (described by a Brenden-Bormann model\cite{rakic98}) and $TiO_2$ as insulator (described by a Cauchy formula given by \cite{devore51}), we have plotted effective index resulting of the 1-MA on Fig.~\ref{dispersionTiO2}. Even if the system width is very small we can see 1-MA predictions are in excellent agreement with the exacts results, at least in the visible range ($0.19 \leqslant \frac{\omega}{\omega_p} \leqslant 0.38$).

Previous works have shown that a drastic impact of nonlocality could be expected when the losses are artificially decreased: the bend-back that is predicted by Drude's model simply disappears when nonlocality is taken into account\cite{raza13}. The example we have chosen shows (see Fig. \ref{dispersionTiO2}b) that burying the metallic slab in a high index dielectric produces the same effect. By moving the frequency of the bend-back away from the interband transitions, it decreases the losses and allows this dramatic change to occur on a very realistic case. This shows that, generally, high index dielectrics have the potential to increase the impact of the spatial dispersion.


Now, we determine above which thickness $h_{lim}$ the one-mode approximation can be fully trusted. 

We have considered here the SRSP when the metallic slab is embedded in $Ti0_2$. This constitutes the ``worst case scenario'' because (i) $TiO_2$ is the transparent material with the highest optical index available in the visible and UV range\cite{xu13}  and (ii) the SRSP mode has the highest possible $k_x$ and thus the highest sensitivity to nonlocality. We have then arbitrarily chosen to assess the accuracy of the 1-MA by comparing the quantities $R_{exact}=\overline{\overline{r}}_{m1}\overline{r}_{m2}$ and $R_{1MA}=r_{m1}r_{m2}$ because they appear in the dispersion relation of the guided mode, whether the exact or approximated relation is considered. Finally, we define $h_{lim}$ as the thickness above which the relative error made on computing $R_{exact}$ using $R_{1MA}$ instead is smaller than $10^{-4}$.

\begin{figure}\centering
\scalebox{1}{\input{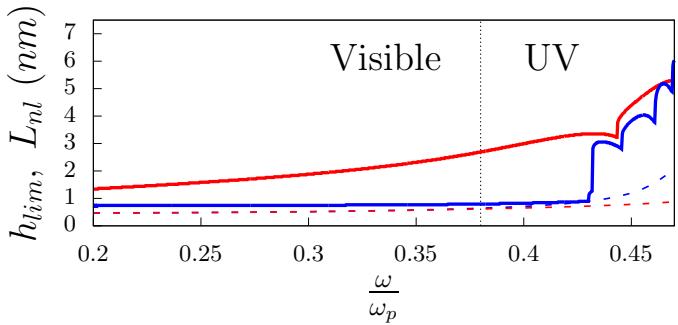}}\label{hlimrutile}
\caption{$h_{lim}$ and penetration depth $L_{nl}$ as a function of the normalized frequency $\frac{\omega}{\omega_p}$ for an IMI slab using Ag as metal and $TiO_2$ as insulator. Red sold lines and dotted lines correspond respectively to $h_{lim}$ and $L_{nl}$ for the SRSP. Blue solid lines and dotted lines correspond respectively to $h_{lim}$ and $L_{nl}$ for the LRSP.}
\label{hlim}
\end{figure}

Given this definition, $h_{lim}$ actually depends on the frequency that is considered. Fig.~\ref{hlim} thus shows the longitudinal wave penetration depth $L_{nl}$ and $h_{lim}$ for visible and near UV range. Globally, $h_{lim}$ behaves similarly to $L_{nl}$ when $\omega$ changes - which is a sign that the 1MA is perfectly accurate when $h$ is in fact large enough compared to $L_{nl}$. The relation between the two is however not straightforward, as the graph shows.

We underline here that we have chosen strict criteria for the 1MA to be deemed accurate. Fig. \ref{hlim} shows what would be $h_{lim}$ for the LRSP too. While in the visible a thickness larger than typically 2.7 nm is required, for the much more studied LRSP, a thickness larger than 1 nm is sufficient. 
Figure \ref{dispersionTiO2}, shows for a $1.7$ nm slab  that even when $h<h_{lim}$, the 1-MA can be considered very accurate in describing a dramatic change in the dispersion relation compared to the local case. 

The values we give for $h_{lim}$ on Fig. \ref{hlim} can thus be fully trusted. They could be discussed only when considering both a material with a higher index than rutile ($TiO_2$) and modes with a higher effective index than the SRSP. But we have checked that even for multilayers there is no increase of this limit, and we underline that in much more realistic situations with lower index dielectrics, like air or glass, and lower effective index (typically smaller than 10) than our worst case considered here, the limit for which the one-mode approximation becomes fully valid is around 1 nm. For a given $\omega$, if the thickness of a metallic slab is larger than $h_{lim}$ there is thus absolutely no need to go beyond the one mode approximation, and the convenient analytic expressions we have derived above are fully sufficient.


In a recent work\cite{luo13}, it has been shown that a well chosen dielectric layer in terms of permittivity and thickness can be designed to fit the reflection coefficient of a bare metallic surface when nonlocality is taken into account. The present work can be considered as a more thorough justification of their work, showing the approach by Luo {\em et al.} can be considered effective for multilayers in general. We have actually found that adding a thin layer of a dielectric with a well chosen permittivity at each interface between a metal and a dielectric allows to match all the Fresnel coefficients we have introduced above, not just the reflection coefficient between a dielectric and a metal. This can be done provided the dielectric layer of thickness $\Delta d$ added at any interface between a metal and a dielectric presents a permittivity 
\begin{align}
\epsilon_t = -\Delta d\,\frac{k_{x}^2}{\Omega}.
\end{align}
We underline that this simple expression is a generalization of the expression given in \cite{luo13} because it allows to change the boundary conditions that are considered by simply changing $\Omega$ and $\kappa_l$\cite{moreau13}.

For multilayered structures, though, the simplification we propose here is even simpler than adding a supplementary layer -- but we admit it cannot be directly extended to complex geometries, whereas adding a dielectric layer is always both possible and easy. And as long as the nonlocal Fresnel coefficients are somehow reproduced, then the impact of nonlocality is likely to be accurately predicted. However, we underline that transformation optics\cite{ward96,Pendry2006}, which has been used to link the response of a multilayer to geometries like metallic edges\cite{luo10} or sharp metal protrusion\cite{luo11} and can take nonlocality into account\cite{fernandez2012transformation,pendry2012transformation}, can be combined with our one-mode approximation. This is expected to simplify such calculations when studying nonlocality and to allow our conclusions to be extended to other geometries.


In conclusion, we have proposed here analytic expressions for the reflection coefficient of a single nonlocal metallic slab using a generalized Fabry-Perot formula, and a simplification of this expression that can be applied to any metallic layer larger than 2.7 nm in the visible range for any wavelength and any surrounding dielectrics. This allows us to introduce nonlocal Fresnel coefficients thus simplifying all the analytic calculations that can be made for metallo-dielectric multilayers. In this framework, we underline that our work is a rigorous justification of why adding a very thin dielectric layer with a well chosen permittivity\cite{luo13} can yield very accurate results.
While exploring different cases to test the robustness of our approach, it appeared that surrounding metallic films with high-index dielectrics is likely to enhance nonlocal effects by allowing to excite wavevector guided modes at lower frequencies for which absorption is lowered. Exciting such guided modes may be experimentally challenging, but it is worth looking in that direction to come up with well controlled experiments showing nonlocal effects.
We hope this work will help the community to assess much more easily the influence of nonlocality on the response of metallic structures with nanometer-sized features - especially in cases that are both numerically expensive and the most likely to be influenced by spatial dispersion\cite{moreau13}, like the computation of the Purcell effect\cite{akselrod14,faggiani2015quenching} when emitters are placed under an optical patch antenna\cite{moreau12b}. Our study actually suggests that, when using advanced simulation tools for complicated geometries\cite{toscano2012modified,schmitt2016dgtd}, it is not necessary to use the hydrodynamic model to describe the response of the metal beyond a boundary layer of 2.7 nm in the visible. This is likely to make such simulations less expensive, an obvious need\cite{luo13} of the community.


\section*{Acknowledgments}
This work has been supported by the French National Research 
Agency, "Physics of Gap-Plasmons" project number ANR-13- 
JS10-0003.

\bibliography{nonlocal}

\end{document}